\documentclass[10pt, conference]{IEEEtran}

\usepackage{hyperref}
\usepackage{romannum}
\usepackage{graphicx}
\usepackage{amsmath}
\usepackage{amssymb}
\usepackage{booktabs,threeparttable}
\usepackage{footnote}
\makesavenoteenv{tabular}

\begin{document}
\begin{titlepage}
    \copyright \, 2020 IEEE.  Personal use of this material is permitted.  Permission from IEEE must be obtained for all other uses, in any current or future media, including reprinting/republishing this material for advertising or promotional purposes, creating new collective works, for resale or redistribution to servers or lists, or reuse of any copyrighted component of this work in other works.
\end{titlepage}

\title{A Comparison of Machine Learning Algorithms Applied to American Legislature Polarization}

\author{\IEEEauthorblockN{Gabriel Mersy\IEEEauthorrefmark{1},
Vincent Santore\IEEEauthorrefmark{2},
Isaac Rand\IEEEauthorrefmark{3}, 
Corrine Kleinman\IEEEauthorrefmark{4}, \\
Grant Wilson\IEEEauthorrefmark{5},
Jason Bonsall\IEEEauthorrefmark{6},
Tyler Edwards\IEEEauthorrefmark{7}
}
\\
\IEEEauthorblockA{\IEEEauthorrefmark{1}University of Minnesota, Minneapolis, USA\\ 
Email: \href{mailto:mersy006@umn.edu
 }{\texttt{mersy006@umn.edu}} }
\IEEEauthorblockA{\IEEEauthorrefmark{2}The Burr Project, Voorhees, USA\\ 
Email: \href{mailto:vince@theburrproject.com
 }{\texttt{vince@theburrproject.com}}} 
\IEEEauthorblockA{\IEEEauthorrefmark{3}University of Chicago, Chicago, USA}
\IEEEauthorblockA{\IEEEauthorrefmark{4}Georgetown University, Washington D.C., USA}
\IEEEauthorblockA{\IEEEauthorrefmark{5}Tulane University, New Orleans, USA}
\IEEEauthorblockA{\IEEEauthorrefmark{6}University of Nevada, Reno, USA}
\IEEEauthorblockA{\IEEEauthorrefmark{7}Temple University, Philadelphia, USA}
}

\maketitle

\begin{abstract}
We present a novel approach to the measurement of American state legislature polarization with an experimental comparison of three different machine learning algorithms. Our approach strictly relies on public data sources and open source software. The results suggest that artificial neural network  regression has the best outcome compared to both support vector machine and ordinary least squares regression in the prediction of both state House and state Senate legislature polarization. In addition to the technical outcomes of our study, broader implications are assessed as a means of highlighting the importance of accessible information for the higher purpose of promoting civic responsibility. 
\end{abstract}


\IEEEpeerreviewmaketitle

\section{Introduction}
The preventative cost of relying on exhaustive research surveys to analyze large-scale social phenomena is irrefutably high. With the increasing prevalence of accessible data sources, predictive models are being leveraged to estimate and categorize data with a higher degree of accuracy than ever before. Machine learning is a field that has been adapted to many different industries such as health care, financial services, and retail, as well as to many domains in academia.

In 2014, Pew Research conducted a survey that aimed at characterizing the degree to which the American public has deviated from the ideological center in recent years. The survey shows that, in the general population as time progresses, both self-identified Democrats and Republicans are falling further away from the ideological center \cite{pew_figure}. In 2006, Abramowitz introduced evidence that supports the argument of polarized political division by highlighting factors such as religiosity, geography, and education \cite[p.~542]{Abramowitz}. We will show that Abramowitz's theory of divergence among the masses is supported by the association between legislative polarization and a wide variety of social, geographic, cultural, and technological factors.   

In parallel with the increasing levels of polarization within the populace, a similar trend in American legislature polarization has been observed \cite{barber_mccarty_2015}. A comprehensive study of American legislatures conducted in 2011 by Boris Shor and Nolan McCarty established a quantified mapping of legislature polarization at the state level \cite[p.~530]{shor_mccarty_paper}. The scope of the ensuing paper is to apply machine learning methods to easily-accessible data sources in order to accurately predict the Shor and McCarty state legislature polarization metrics. In a broad sense, we intend to highlight the predictive power of surveyed demographic data when applied to sociopolitical phenomena. In addition to validating the predictive ability of our model, we aim to depict the relationship between demographic features and legislative polarization.

The paper is structured as follows: we first provide commentary on related works that informed our approach (Section \Romannum{2}). We then outline a framework for the aggregation and exploratory analysis of social data sources to select domain-appropriate features (Section \Romannum{3} and \Romannum{4}, respectively). We proceed to fit three machine learning models to the 2013-2016 data as well as assess their performance using cross-validation and hypothesis testing (Section \Romannum{5}). In the same section, we comment on the practicality of using fitted models to predict future years that fall outside of the original time horizon of the data. To highlight this, we predict the state House and Senate polarization metrics for the year 2017 with the best-performing model (ANN). We conclude with the implications of our work at scale, which includes relevant qualitative discussion concerning the relationship between technology, information availability, and civic responsibility (Section \Romannum{6}). 

\section{Related Work}
Now turning to the problem at hand, there is a wide body of research concerning machine learning in the social sciences. The emerging research field of \textit{computational social science} is an interdisciplinary approach that exists at the intersection of the social sciences, statistics, and computer science \cite{mason_winter_2014}.  In particular, political scientists have primarily concerned themselves with causal inference as opposed to descriptive statistics. Notably, in 2012, John Gerring showed that approximately 80 percent of articles in the \textit{American Political Science Review} concerned causal inference \cite{gerring_2012}. Therefore, generally speaking, political science problems present as ideal opportunities to employ the use of machine learning algorithms \cite{grimmer_2015}. 

The technical literature aimed at leveraging machine learning and exploratory data analysis on political science problems is still in its infancy. Conover et al. introduced the problem of political polarization analysis by constructing network topologies for political retweets and direct mentions via unsupervised machine learning methods \cite{conover_2011}. Garcia et al. presented a case study that constructs analyses to expose the mechanisms facilitating the diffusion of political information on YouTube \cite{garcia_2012}. Khaze et al. provided a specific approach that employs the use of an artificial neural network to predict voter turnout in Iran \cite{Khaze_2013}. Lastly, Grundler implemented a SVM to measure democracy \cite{GRUNDLER_2016}. 

\section{Data Aggregation}
\subsection{The Shor-McCarty Data}
American political coverage frequently focuses on the national levels of operation: legislature in Congress, Executive action in the White House, and rulings of the Supreme Court. This attention at the national level ignores the complex trends that make up the state and local politics that are the foundations of how American people interface with their government. For a democracy to thrive, civic engagement is essential---not only at the national level, but at the state and local levels as well.

The Shor-McCarty data set examines how state-level political polarization affects good governance and influences the direction of public policy. The data set contains state-level legislature ideology metrics by chamber from 1993 to 2016 including the margin of partisan majority, average ideology difference of the chamber’s members, and the difference in median partisan ideology. Examining state-level data in its own right and pairing it with US Congressional data gives analyses a greater depth, allowing for more profound insight into the policy decisions and state issues that result in our polarized national discourse. The results reflect how state party lines do not necessarily correlate uniformly to their respective national party. They leave their data with a call to action, hoping that their data set becomes the foundation of future work on state level legislative data \cite[p.~550]{shor_mccarty_paper}. The labels of state House polarization and state Senate polarization were collected from the Shor-McCarty database \cite{dataset_main}. Each polarization metric is a continuous positive number ($\mathbb{R}^{+}$). We only focused on the most recent years (2013-2016).

\subsection{The Data Sets}
The features were manually collected from four primary sources. The first and largest source is the American Community Survey (ACS) maintained by the U.S. Census Bureau. Numerous demographic features for the years of 2013 through 2017 were compiled \cite{acs}. The second source was the Stanford Mass Shootings in America database \cite{msa}. To parse the MSA data into the correct format for the supervised learning task, a Python script was written to aggregate the data by state and year. A recent Yale study that surveyed everyday Americans to assess their attitudes on climate change was included \cite{climate}. The same aggregation approach to that of the MSA data set was employed. The final source was a collection of surveys concerning religiosity in America by state and year \cite{religiosity}. 

\subsection{Arrangement for Supervised Learning}
Before arranging the data to obtain the format necessary for the supervised learning tasks, the \textit{feature} samples from 2013 to 2016 were separated from the year 2017, due to the fact that 2017 is outside of the scope of the original Shor-McCarty study. Without corresponding label values, the 2017 features were stored to predict the two polarization metrics for each state using the best-performing models (Section \Romannum{5}). 

The features and labels were then pooled into two separate CSV files: one for the House polarization regression task and one for the Senate polarization regression task. The complete data set contains a total of 180 complete samples for the House polarization set and an equivalent 180 samples for the Senate polarization set. A total of 15 features were included prior to the feature selection process such that any given complete feature sample has a corresponding label value for both legislative bodies. The data is ordered by two characteristics: increasing year and alphabetical order by state name.  The representation for the supervised learning problems is given below with $m = 180$ rows and $n = 15$ columns. Each row represents a state, $h_i$ and $s_i$ are the House and the Senate polarization indices from \cite{dataset_main}, and $x_{ij}$ is the pointwise value of the feature $X_j$ for the state corresponding to the index $i$.

 \[ \left\{\begin{bmatrix}
           \mathbf{h_{1}} \\
           h_{2} \\
           \vdots \\
           h_{180}
         \end{bmatrix}, %
         \begin{bmatrix}
           \mathbf{s_{1}} \\
           s_{2} \\
           \vdots \\
           s_{180}
         \end{bmatrix},
         \begin{bmatrix} %
\mathbf{x_{11}}&\mathbf{x_{12}}& \mathbf{\cdots} &\mathbf{x_{1n}} \\
x_{21}&x_{22}&\cdots &x_{2n} \\
\vdots & \vdots & \ddots & \vdots\\
x_{n1}&x_{n2}&\cdots &x_{mn}
\end{bmatrix} \right\}\]

For example, consider the boldface first row of the arrays within the set given above. Specifically, $h_1$ and $s_1$ are Alabama's 2013 Shor-McCarty polarization metrics for its House and Senate legislatures. These values are $0.73$ and $0.93$ respectively. The data point $x_{11}$ is Alabama's percent of people in 2013 who had an internet subscription, an estimated $64.7$ percent. Similarly, the data point $x_{19}$ is Alabama's median age from 2013, $38.3$. Note that $h_{180}$, $s_{180}$, etc. correspond to Wyoming's metrics for 2016.

\section{Feature Selection}
In the prediction of political polarization, some degree of domain interpretation of the relationships between variables is markedly important. In addition to interpretation, the prediction accuracy of black box methods is particularly attractive. Both of these aspects are desirable when disseminating results that have broader ethical implications. Although Pearson's coefficient $r$ was used to select the most appropriate features, we introduce a novel data mining algorithm that is reliant upon the Pearson and Spearman correlation coefficients to flag potential non-linear relationships. 

Pearson’s correlation coefficient $r$ measures the direction and strength of a linear relationship. In contrast to Pearson's coefficient, Spearman's coefficient $\rho$ is a non-parametric test of the strength and direction of an increasing or decreasing relationship between two ranked variables. Thus, continuous variables are reduced to a ranking of their values. In contrast to Pearson's coefficient, $\rho$ captures non-linearities, since it does not assume the underlying distribution and is more resistant to outliers \cite{spearman}.

\subsection{Non-Linearity Flagging Algorithm}
In this section, we present a novel algorithm for flagging potential non-linear relationships between a feature set and a given label. When one encounters an extensive feature set where the underlying relationships are largely unknown, correlation coefficients and scatter plots tend to be the first step in exploratory data analysis. The purpose of this algorithm is to extract useful information about the features potentially indicating non-linear relationships with the label as well as provide insight into the existence of extreme outliers within a particular feature.

Let $X = \{X_1, X_2, ..., X_m\}$ be a set of feature vectors, and let $y$ be the corresponding label vector. Suppose that the functions $r(X_j, y)$, $\rho(X_j, y)$ are each defined to return a value $r, \rho \in [-1, 1]$ indicating the respective correlation coefficient between feature $X_j$ and label $y$. The algorithm flags features where $\rho - r \ge \lambda$ for a specified tolerance $\lambda$. The decision boundary is represented as  
\begin{equation}
    1- {\frac {6 \sum_{i=1}^{n} d_i^2}{n(n^2 - 1)}} - \frac{{}\sum_{i=1}^{n} (x_i - \overline{x})(y_i - \overline{y})}
{\sqrt{\sum_{i=1}^{n} (x_i - \overline{x})^2(y_i - \overline{y})^2}}  \ge \lambda
\end{equation}
\noindent where $d_i$ is the difference between the discrete ranks of an observation, $x$ is the sample feature, $y$ is the sample label, and $\bar{x}$ and $ \bar{y}$ are the means of the sample feature and the sample label respectively.

Relying on domain knowledge concerning the comparatively greater extent of limitation surrounding the measurement of social metrics, we chose a relatively low $\lambda = 0.05$ for our data set. Our rationale behind choosing a small tolerance in this case is centered around the fallibility of measurement within the social sciences. One notable limitation highlights the prevalence of \textit{habitual measurement practices} that have little regard for the evolving methodological advancements aimed at improving the validity of social knowledge representations \cite[p.~16]{social_science_correlation}. We decided that a small tolerance was appropriate for our data set due to the intrinsic relationship between fallibility of measurement and statistical correlation \cite[p.~15]{social_science_correlation}. However, these empirical results do not discount the practicality of employing our algorithm to analyze extensive data sets. In total, 5 features were flagged in the House data set, and 4 features were flagged in the Senate data set. None of these flagged features appeared in the final feature set. 

The results of applying the algorithm suggest that the reason these features were flagged was due to the existence of outliers. This conclusion was reached after an examination of a pair plot between these features and the corresponding labels of interest. Furthermore, the results could indicate that the features chosen are mostly linearly-related to the labels in the aggregated polarization data sets. An empirical assessment of the problem domain is often the best manner of choosing a tolerance $\lambda$. In particular, this conclusion allows us to deduce that the p-values corresponding to the coefficients in an OLS regression may provide us with a reasonable depiction of the trends underlying feature importance as described in \cite{rudin_2019}.

\subsection{Final Features}
To obtain the final set of features, Pearson's $r$ values were computed. The following set of features with $r \ge 0.1$ \cite{cohen_1998} (with the exception of two of the House features) were passed to the modeling stage.\footnote{An identical set of 9 features were included in the final feature set for both tasks to maintain consistency. The incremental increase in RMSE resulting from the inclusion of the two features in the House task with $r < 0.1$ was negligible for each model ($< 0.02$).} The 9 features included in each model are given below along with their corresponding correlation pairs in the format $(r_h, r_s)$ ordered by strength of association, where $r_h, \: r_s$ represent the correlation between that feature and the specified label (see Fig. 1). 

\begin{enumerate}
    \setlength{\itemsep}{0pt}
    \item Total number of people who moved to a particular state in the United States from abroad  $(0.383, 0.463)$ 
    \item Percent of the population who believe in climate change $(0.266, 0.352)$
    \item The number of mass shootings defined as 3 or more victims $(0.223, 0.243)$
    \item Percent of the civilian non-institutionalized population with a disability $(-0.204, -0.244)$
    \item Percent of the population who considered themselves very religious $(-0.207, -0.2111)$
    \item Percent of the population 25 and greater years of age with a bachelor’s degree $(0.197, 0.226)$
    \item Percent of the population who considered themselves non-religious $(0.177, 0.194)$
    \item Household median income $(0.071, 0.163)$
    \item Percent of the population with an internet subscription $(0.053, 0.105)$
\end{enumerate}

\begin{figure}[h]
\centering
\includegraphics[scale = 0.4]{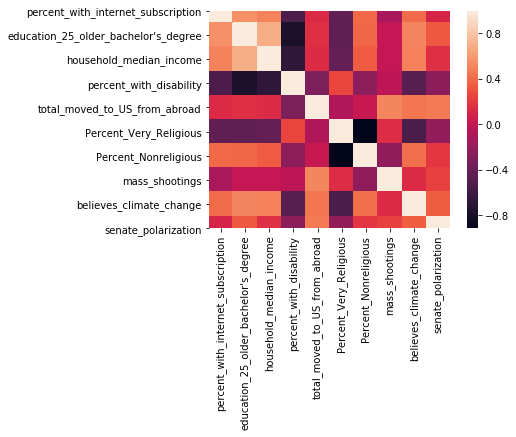}
\caption{Heat map of Pearson correlations between the select features with $r \ge 0.1$ and the Senate polarization label} 
\end{figure}

\section{Machine Learning}
To build the machine learning models for the regression task of predicting state legislature polarization, we employed the use of two Python libraries: scikit-learn for OLS and SVM regression, as well as Keras (TensorFlow back-end) for the ANN \cite{sklearn}, \cite{keras}.\footnote{The code for our machine learning methodology can be found \href{https://github.com/the-burr-project/A-Comparison-of-Machine-Learning-Algorithms-Applied-to-American-Legislature-Polarization}{\texttt{here}}.}  

\subsection{Data Preparation}
Before implementing each model, a randomized validation split partitioned the 2013-2016 data set into training and validation subsets. A total of 144 samples (80 percent) were used for model training. A total of 36 samples (20 percent) were used for model validation. Following this process, min-max scaling was applied to map the data points from the training and validation sets into the reduced range $[0,1]$ to improve algorithm convergence speeds.  

\subsection{Performance Metric}
Root Mean Square Error (RMSE) is employed to compare performance across models \cite{RMSE}. Thus, we define our loss function $\mathcal{L}(y, \hat{y})$ as the RMSE where $\hat{y}$ is the vector of predictions from the model and $y$ is the vector of true values.  
\begin{equation}
\mathcal{L}(y, \hat{y}) = \sqrt{\frac{1}{n} \sum_{i=1}^{n}{(\hat{y}_i -y_i})^2} = \mathrm{RMSE}
\end{equation}

\noindent We choose the model $m$ yielding the set of predictions $\hat{y}^{*}$ minimizing the validation loss $\mathcal{L}(y, \hat{y})$ such that
\begin{equation}
\hat{y}^{*} = \underset{\hat{y} \in P}{\mathrm{argmin}} \; \mathcal{L}(y, \hat{y})
\end{equation}
where $P$ is the set of predictions resulting from each model. 

\subsection{Ordinary Least Squares Regression}
OLS regression seeks to minimize the residual sum of squares given a matrix of features $X$ and a label vector $Y$. OLS regression is included as a heuristic performance benchmark for our comparative analysis. 

\subsection{Support Vector Machine}
In support vector machine (SVM) regression, we use hyperplanes to construct a function $\hat{f}(X)$ that will estimate our target $Y$ where $Y - \hat{f}(X) < \epsilon$ for some maximal margin $\epsilon$. We rely on a kernel to efficiently evaluate the linear transformation $\phi: \mathbb{R}^n \to F$ which maps the input space $\mathbb{R}^n$ to the higher-dimensional feature space $F$. Our implemented models use a radial basis function kernel (RBF). Our SVM regression model is given by 
\begin{equation}
    \hat{f}(X) = \sum_{i=1}^{l} v_i \cdot \exp \left[\frac{-||x-x'||^2}{\sigma^2} \right] + b
\end{equation}
where the parameters $v_i \in V$ are the solutions to a constrained optimization problem, $||x-x'||$ refers to the Euclidean distance between the feature vectors of two unique samples, $\sigma$ constrains the linearity of the decision boundary in $F$, and $b$ is the bias term \cite{hearst}. After model building, hyperparameter optimization was carried out by experimentally evaluating different combinations of the model's parameters. 

\subsection{Artificial Neural Network}
The fundamental unit for a feedforward neural network is the neuron. The output tensor $y$ for each neuron with respect to its input tensors $x_i \in \{x_1, x_2, ..., x_n\}$ is given by
\begin{equation}
    y = g\left(\sum_{i=1}^{n}w_{i}x_{i} + b\right)
\end{equation}
where the parameters $w_i \in \{w_1, w_2, ..., w_n\}$ are the weights multiplied by each input $x_i$, $b$ is the bias parameter, and $g(\cdot)$ is a pointwise activation function. A feedforward neural network consists of an arrangement of neurons structured by an input layer, various hidden layers, and an output layer. For our ANNs, hyperparameter optimization was carried out in a similar approach to that of the SVM model. As a result, the ANN model for each polarization prediction task is composed of an input layer of 9 neurons, a single hidden layer of 5 neurons, as well as a dropout layer of  $p = 0.2$. The stochastic optimization algorithm was set as \textit{Adam} with default parameter values as specified by \cite{keras}. Pointwise rectified linear unit activation (ReLU) was applied to each neuron in the hidden layer such that $g(\cdot) = \max(0, \cdot)$.

\subsection{Results}
RMSE values on the validation set of $v = 36$ samples for each of the 3 models are given in Table \Romannum{1}. These values are split up by the House and Senate polarization prediction tasks. A two-tailed Wilcoxon non-parametric signed-rank test was carried out to determine whether a difference exists between the underlying sampling distributions of the model predictions at the significance level $\alpha = 0.05$ \cite{wilcox}. The results of the tests are given in Table \Romannum{2}.

\begin{threeparttable}[htb]
    \caption{validation set results $v = 36$ samples}
    \small
    \setlength\tabcolsep{0pt}
\begin{tabular*}{\linewidth}{@{\extracolsep{\fill}} c c c }
    \toprule
   Model & House RMSE & Senate RMSE \\
   \cmidrule{1-3}
OLS          & 0.421  & 0.328 \\
SVM       & 0.386  & 0.308  \\
ANN       & \textbf{0.372}\tnote{†}  & \textbf{0.289}\tnote{†}  \\
        \bottomrule
\end{tabular*}
\begin{tablenotes}
            \item[†]Best performing model $m := \hat{y}^{*}$ 
\end{tablenotes}
\end{threeparttable}

\begin{threeparttable}[htb]
    \caption{Wilcoxon signed-rank test on validation set predictions}
    \small
    \setlength\tabcolsep{0pt}
\begin{tabular*}{\linewidth}{@{\extracolsep{\fill}} c cc cc}
    \toprule
   \multicolumn{1}{c}{Models} & \multicolumn{2}{c}{House} &  \multicolumn{2}{c}{Senate} \\
   \cmidrule{1-5}
    & $W$ &  p-value & $W$ &  p-value \\

OLS-SVM &  90 & $1.34 \times 10^{-4}$\tnote{§} & 321 & 0.850 \\
OLS-ANN & 116 & $6.51 \times 10^{-4}$\tnote{§} & 79 & $6.59 \times 10^{-5}$\tnote{§} \\ 
SVM-ANN & 303 & 0.637 & 29 & $1.79 \times 10^{-6}$\tnote{§}  \\
        \bottomrule
\end{tabular*}

\begin{tablenotes}
            \item[§]Statistically significant at $\alpha = 0.05$
\end{tablenotes}

\end{threeparttable}

\begin{figure}[h]
\centering
\includegraphics[scale = 0.47]{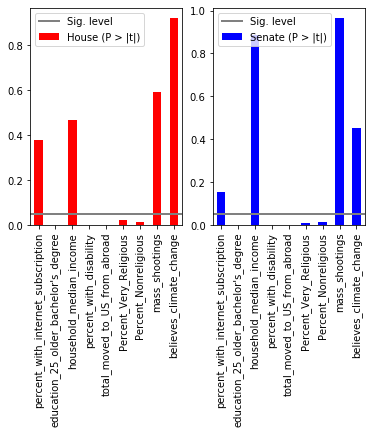}
\caption{P-values for the OLS regression coefficients}
\end{figure}

\subsection{Discussion}
ANN regression performed the best when comparing its loss metric to SVM and OLS regression for each task. At the epochs where the two final models were saved, the networks had a similar, but slightly lower training loss when compared to the validation loss. This implies that the model at the chosen epoch did not overfit the data. The ANN model had a lower loss for the Senate polarization data set than the House polarization on average as well as a longer convergence time when compared to the House ANN. One potential reason for this is the presence of markedly lower correlation values between the features and the House label (Section \Romannum{4}). 

The Wilcoxon hypothesis test on the predictions from the ANN-SVM pair suggests that the two models generated similar sampling distributions for the \textit{House task}. Furthermore, a juxtaposition of the House OLS-SVM and the OLS-ANN pairs shows that each model yields a unique sampling distribution. In the case of the \textit{Senate task}, we observe an occurrence that is distinctive from the House task. In the significance test between the predictions for the ANN-SVM and OLS-ANN pairs, the models generate unique sampling distributions, while in the OLS-SVM pair, the models generate similar distributions. Additionally, the conclusions from the non-linearity flagging algorithm (Section \Romannum{4}) suggest that OLS regression may allow us to obtain a reasonable depiction of feature importance trends for the SVM and ANN models by analyzing the p-values for each coefficient (Fig. 2).\footnote{These are \textit{not} the same coefficients as the original model---this version does not include the y-intercept, $b$, so that we can focus on the features themselves.} A limitation to this interpretation approach is that we assume that each feature and label is distributed normally, which is not necessarily the case. As such, the results in Fig. 2 should not be seen as an absolute truth. 

After deciding upon the two ANN models by evaluating their comparative performance on the validation set, predictions were made for a 2017 testing set of features for which there is no polarization label\textemdash the data for the 2017 polarization metrics do not exist. This new task was oriented on extrapolating the Shor-McCarty results to display the domain novelty of our two models. The geographical depiction of these new 2017 values for the House and Senate are included in Fig. 3 and Fig. 4, respectively. 

\begin{figure}[h]
\centering
\includegraphics[scale = 0.27]{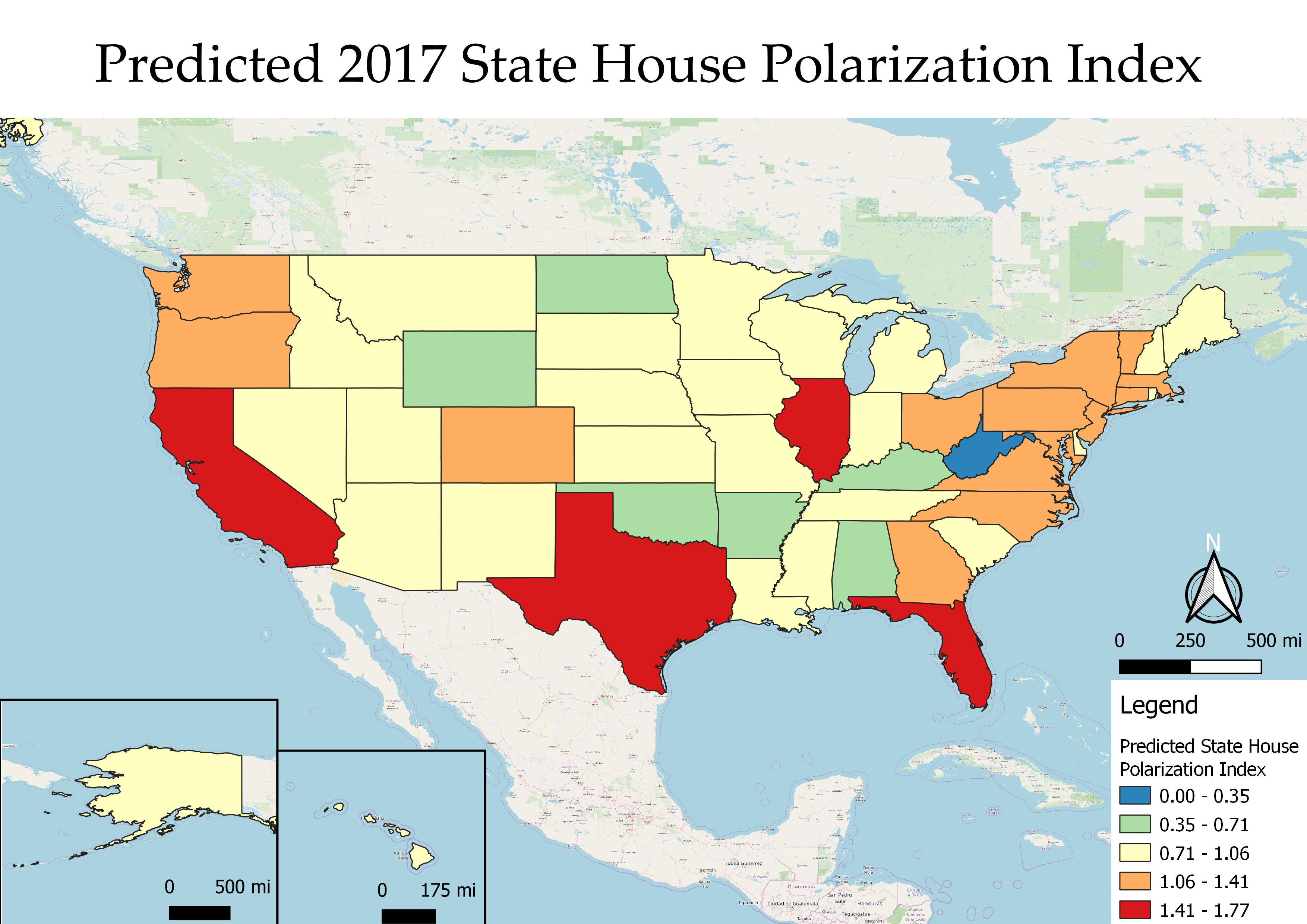}
\caption{2017 House Predictions} 
\end{figure}

\begin{figure}[h]
\centering
\includegraphics[scale = 0.27]{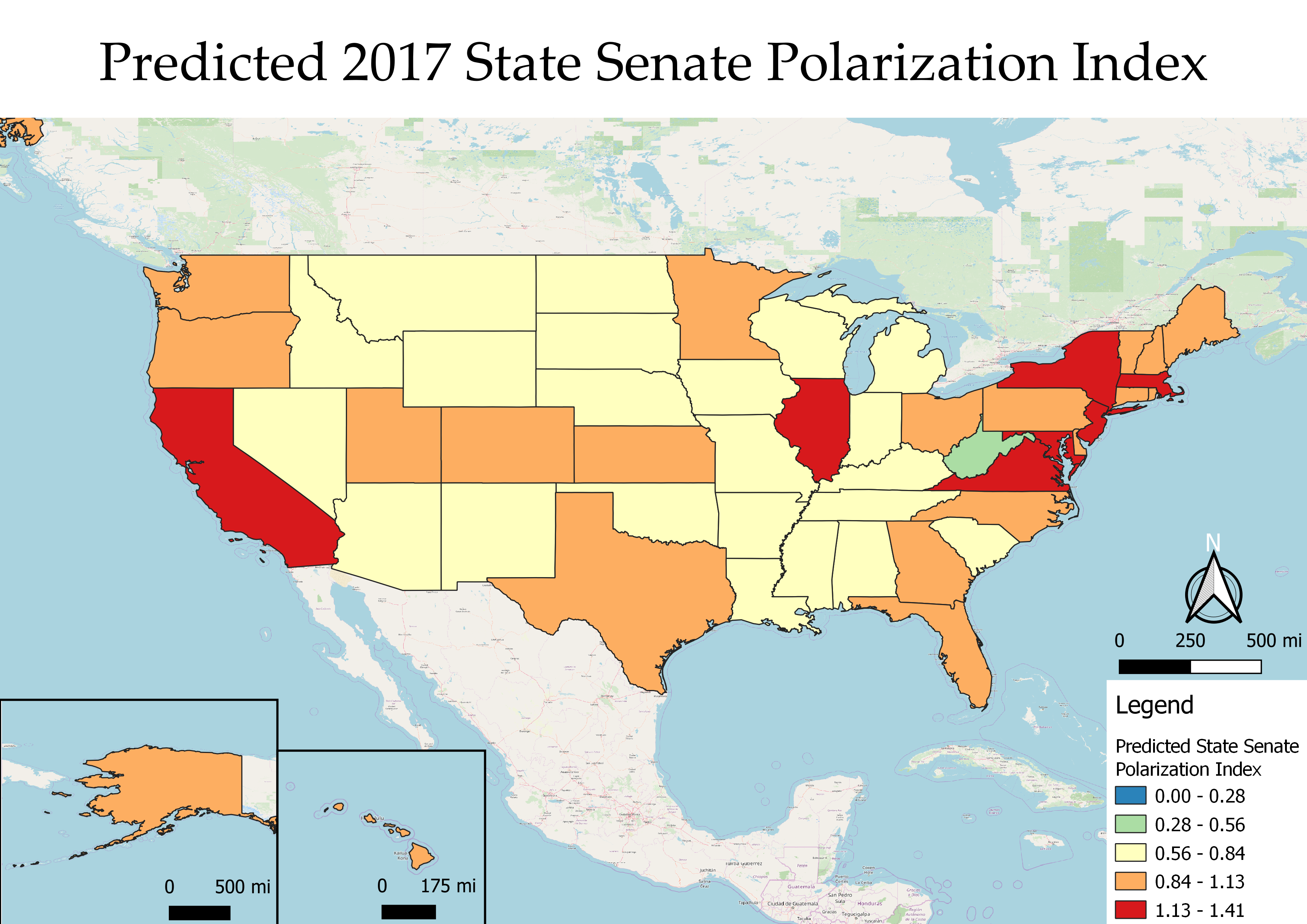}
\caption{2017 Senate Predictions} 
\end{figure}

\section{Conclusion}
We introduced a novel approach to the measurement of state Legislature polarization using machine learning. From the technical results, we conclude that ANN regression performed the best in predicting state Legislature polarization relative to SVM and the benchmark model of OLS regression. In addition to this, our results suggest that the features included in our models are appropriate for the measurement of state Legislature polarization. With the application of machine learning to this particular societal problem, we aimed at implementing a solution to the time delay that is typically associated with generating an up-to-date political science domain ontology. Though the technical results of our study are significant in their own right, we would like to expand upon an argument for the societal importance of our models. 

We the People must come to an understanding of the level of polarization that is dividing our communities, states, and the country as generations have known it. Abramowitz concluded in his 2006 paper that the American people---to no small extent---must hold themselves responsible for the divisions occurring within the political system \cite[p.~543]{Abramowitz}. It is of the utmost importance to understand the consequences of our collective and individual actions in electing representatives who fall towards the extreme ends of the political spectrum. Information availability by means of technological advancement is a fundamental first step towards that of a harmonious movement to oust partisan representatives and unite the system. Thankfully, in the appropriately titled Information Age with digital accessibility amongst Americans and their elected officials at an all-time high, this goal is as achievable as it has ever conceivably been. It is the responsibility of representatives to come together to effectively represent the People. In the same vein, it is the People’s responsibility to hold their representatives to the task of overcoming partisan division. 

Hence, on a qualitative level we conclude that (1) through the introduction of big data and machine learning into the political process, we now have the computing power to provide timely information to the People. Of utmost importance, (2) the availability of information depicting the level of division occurring within the most fundamental levels of government largely drives the People's responsibility to effect change. However, (3) this same information can soon become obsolete if not for predictive models, statistical surveys, and other methods to model reality with a reasonable degree of certainty. We encourage the research community to build upon our results to continuously update polarization metrics when new data become available. With this in mind, we believe herein lies a consequential first step forward towards the ideal of a truly United State.

\section*{Acknowledgment}
We would like to thank Dr. Maria Gini (University of Minnesota Department of Computer Science and Engineering) and Dr. Anthony Breitzman (Rowan University College of Sciences and Mathematics) for their guidance and mentorship---as well as former Vice President Aaron Burr Jr. for inspiration by pioneering the use of political data analytics during the election of 1800. 

\bibliographystyle{IEEEtran}
\bibliography{IEEEabrv, references.bib}

\end{document}